\documentclass[11pt,a4paper]{article}
\usepackage{iftex}
\ifPDFTeX
\usepackage{mathptmx,courier}
\else
\usepackage{mathptmx}
\usepackage{fontspec}
\IfFontExistsTF{Times New Roman}{\setmainfont{Times New Roman}}{\setmainfont{TeX Gyre Termes}}
\fi
\ifPDFTeX\usepackage[utf8]{inputenc}
\usepackage[T1]{fontenc}\fi
\usepackage[a4paper,margin=25mm]{geometry}
\usepackage{amsmath,amssymb,amsthm,bm,graphicx,booktabs,tabularx,longtable,array}
\usepackage{caption,subcaption,float,placeins,setspace,tikz,microtype,needspace,titling,titlesec,xurl}
\titleformat{\section}{\normalfont\normalsize\bfseries}{\thesection}{.7em}{}
\titleformat{\subsection}{\normalfont\normalsize\bfseries}{\thesubsection}{.7em}{}
\titleformat{\subsubsection}{\normalfont\normalsize\bfseries}{\thesubsubsection}{.7em}{}
\titlespacing*{\section}{0pt}{10pt}{4pt}
\titlespacing*{\subsection}{0pt}{8pt}{4pt}
\titlespacing*{\subsubsection}{0pt}{6pt}{3pt}
\floatstyle{plain}
\newfloat{algorithm}{tbp}{loa}
\floatname{algorithm}{Algorithm}
\floatstyle{plain}
\usepackage[numbers,square,sort&compress]{natbib}
\usepackage[hidelinks]{hyperref}
\hypersetup{pdftitle={When do quantum views help? Complementary classical--quantum representations in multi-view fusion},pdfauthor={Azadeh Alavi and colleagues},pdfsubject={Research manuscript}}
\usetikzlibrary{arrows.meta,positioning,shapes.geometric,calc}

\newcommand{\declarationheading}[1]{\par\medskip\noindent\textbf{#1}\par\nobreak}

\title{When do quantum views help? Complementary classical--quantum representations in multi-view fusion}
\author{\normalsize Azadeh Alavi$^{1,*}$, Fatemeh Kouchmeshki$^{2,*}$,\\
\normalsize Hossein Akhoundi$^2$, Yongli Ren$^1$,\\
\normalsize Abdolrahman Alavi$^2$ and Muhammad Usman$^3$}
\date{\small 13 October 2026}
\renewenvironment{abstract}{\par\begingroup\singlespacing\noindent\textbf{Abstract}\par\endgroup\nobreak}{\par\medskip}
\begin{document}
\doublespacing
\begingroup\singlespacing
{\centering
{\fontsize{14}{16}\selectfont\bfseries When do quantum views help? Complementary classical--quantum representations in multi-view fusion\par}
\vspace{1pt}
Azadeh Alavi$^{a,*}$, Fatemeh Kouchmeshki$^{b,*}$,\\
Hossein Akhoundi$^b$, Yongli Ren$^a$,\\
Abdolrahman Alavi$^b$ and Muhammad Usman$^c$
\par}
\noindent\footnotesize $^a$RMIT University, GPO Box 2476, Melbourne VIC 3001, Australia.\\
$^b$Pattern Recognition Pty Ltd, Melbourne, Australia.\\
$^c$Faculty of Information Technology, Monash University, 25 Exhibition Walk,\\
Clayton VIC 3800, Australia.\\
$^*$Corresponding authors: Azadeh Alavi (\href{mailto:azadeh.alavi@rmit.edu.au}{\mbox{azadeh.alavi@rmit.edu.au}});\\
Fatemeh Kouchmeshki (\href{mailto:admin@pr2aid.com}{\mbox{admin@pr2aid.com}}).
\par\endgroup\normalsize

\begin{abstract}
Recent studies have reported promising results from quantum machine learning in drug discovery, healthcare, materials science and cybersecurity. Research includes comparisons with classical machine learning algorithms and the development of sequential hybrid pipelines. We propose that fusion can improve prediction when classical and quantum representations capture complementary predictive relationships. Our classical--quantum multi-view framework preserves the full classical input alongside simulated quantum features. Concatenation and attention combine the views; single-view comparisons, replacement and masking controls assess both views' contributions. Controlled angular--polynomial experiments distinguish targets adequately represented by a single view from those that depend on complementary angular and polynomial relationships. For these mixed targets, fusion improves prediction by combining the relevant feature sets, while matched classical representations recover the same benefit, linking the improvement to their complementary predictive content. Controlled classification tests further show that gains depend on alignment between the chosen representation and the target, and on a suitable fusion rule. Benchmark validation shows improved mean predictive performance over classical baselines on Wine, Wisconsin Diagnostic Breast Cancer, three-class Fashion-MNIST and Steel Plates Faults within the evaluated protocols. Classical replacement and masking controls remain competitive, classical baselines lead on Digits and Covertype, and outcomes with missing or corrupted views depend on the available inputs. These findings support evaluating quantum representations for their complementary predictive contribution and matching the representation and fusion strategy to the target relationships.
\end{abstract}

\begingroup\singlespacing\small
\par\endgroup\medskip
\doublespacing

\section{Introduction}\label{sec:intro}
Recent studies have reported promising results from quantum machine learning in drug--target interaction prediction, breast cancer diagnosis, materials science and intrusion detection \cite{Drug2025,Health2024,Materials2026,Cyber2024}. These findings motivate further investigation of quantum representations and their integration with established learning methods. Related studies examine robustness \cite{West2023Robustness,Khatun2025Distillation,Dowling2026Robustness}, medical image generation \cite{Khatun2025Generation}, compact sensor representations and learning from small fabrication datasets \cite{Wang2025Sensor,Wang2025Ohmic}. Multi-view learning provides a complementary perspective: different representations of the same observation can support joint prediction, including views derived from one input source \cite{Wang2011SingleSourceMVL}. Shared-subspace learning and fusion across view combinations further develop this idea \cite{Lin2023CommonSubspace,Yang2026HMDMV}. A prominent research objective has been to develop quantum models that outperform classical alternatives \cite{Biamonte2017QML,Schuld2022Goal}. However, predictive gains depend on the task, representation and comparison method. Broad benchmarking has found classical learners competitive with, and often stronger than, the quantum models evaluated \cite{Bowles2024Benchmark}, while classical learning can remain effective even for targets generated through quantum processes \cite{Huang2021Data}.

Many hybrid approaches integrate classical and quantum processing within a sequential feature hierarchy \cite{Mari2020Transfer,Henderson2020Quanvolution}. Classical-to-quantum transfer learning passes pretrained classical features to a trainable quantum model \cite{Mari2020Transfer}. Quantum-to-classical processing appears in quanvolutional networks, where quantum operations transform local image patches before a classical network processes their numerical outputs \cite{Henderson2020Quanvolution}. Dressed quantum circuits place trainable classical layers before and after quantum processing, giving a classical--quantum--classical sequence \cite{Mari2020Transfer}. Quantum feature-map methods also combine a quantum representation with classical kernel learning \cite{Schuld2019Feature,Havlicek2019Features}. These approaches establish the value of studying quantum components within a larger learning system.

\begin{figure}[htbp]
\centering\includegraphics[width=\linewidth]{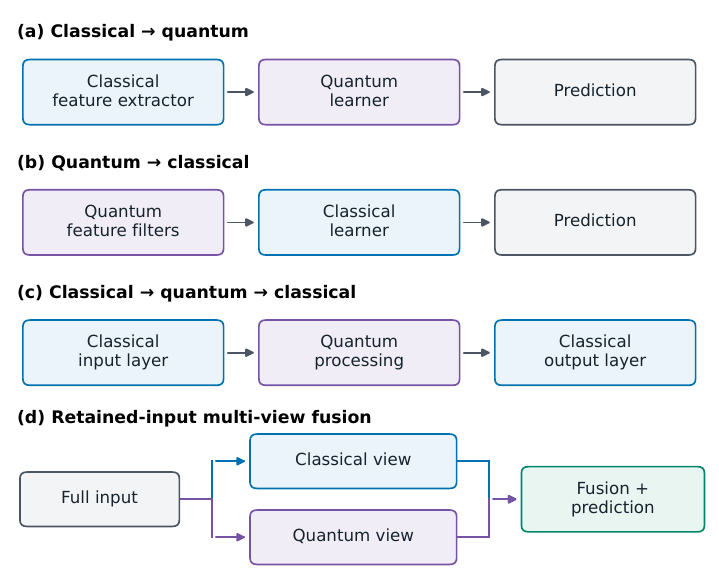}
\caption{Representative hybrid information flows. Panels (a) and (c) summarize classical-to-quantum and classical--quantum--classical arrangements \cite{Mari2020Transfer}; (b) shows quantum feature extraction followed by classical learning \cite{Henderson2020Quanvolution}. Panel (d) shows the retained-input fusion arrangement studied here, with retained-input bypass models as prior architectural precedent \cite{Zhang2025ReadoutBypass_a}. Arrows show forward processing, rather than the training optimizer. Each quantum stage transforms its input into numerical features; Section~\ref{sec:quantum_background} explains the operations involved. The arrangements are schematic and may overlap in a complete model.}
\label{fig:routes}
\end{figure}

Existing hybrids also include retained-input fusion. Retained-input models, described as readout-side bypass in the literature, concatenate the original classical features with quantum-derived numerical features before classification \cite{Zhang2025ReadoutBypass_a}. Our study builds on this architectural precedent and examines complementarity across fusion rules, target constructions and input conditions. We ask how both views contribute to prediction and when their combination improves on either view alone. Replacement and masking controls additionally compare the quantum view with a learned classical alternative and the retained classical path. Figure~\ref{fig:routes} places this question alongside representative sequential arrangements.

The motivation is that a prediction task may depend on several relationships: some are captured effectively by a classical representation, while others may become easier to learn through quantum-derived features. Quantum feature maps provide nonlinear transformations whose usefulness depends on the encoding and the target relationship \cite{Schuld2019Feature,Huang2021Data}. We therefore propose that fusion can improve prediction when the two representations capture complementary relationships relevant to the target. Multi-view learning provides a framework for combining these representations of the same observation \cite{Xu2013Survey,Baltruaitis2018MultimodalReview_a,Wang2011SingleSourceMVL}. Retaining the full classical input allows the fused predictor to use both views, including when circuit features alone do not capture enough structure for accurate prediction. Comparisons with each view alone assess the contributions of both representations. Classical replacement and masking controls further assess whether the quantum-derived features contribute beyond added classical capacity and the fusion mechanism.

We propose a classical--quantum multi-view framework that retains the full classical input as the main predictive path and combines it with features from a simulated quantum circuit, an ordered sequence of quantum operations. Concatenation and attention provide alternative ways to combine the views while preserving direct access to the original features. In the matched experiment, a learned classical replacement supplies the same number of additional features through the same fusion interface, and a masking control prevents the quantum outputs from influencing prediction.

The evaluation comprises four complementary protocols within the same study, beginning with controlled synthetic tasks in which the target relationships are known and then extending to benchmark comparisons. The angular--polynomial diagnostic varies the number of inputs to examine when a single view is sufficient and when fusion helps. Same-input target mixtures and separated feature blocks then test representation alignment, model selection and the fusion rule. A five-benchmark experiment compares fusion stages on Wine, Wisconsin Diagnostic Breast Cancer, Covertype, Fashion-MNIST and Steel Plates Faults. A matched experiment on Wine, breast cancer and Digits introduces the classical replacement and masking controls. Experiments on Multiple Features and an image--text bird benchmark examine an added circuit view when separately available inputs are missing or corrupted. Comparisons remain within each protocol's architecture, data subset and training budget.

The controlled tests examine how the representation matches the target, motivated by quantum feature-map theory \cite{Schuld2019Feature} and studies of engineered learning tasks \cite{Huang2021Data}. The angular--polynomial diagnostic shows that a view suited to angular relationships can be sufficient on its own. When angular and polynomial relationships both determine the target, fusion achieves mean root mean squared errors of 0.1037, 0.1056 and 0.1080 at three, four and five angular inputs, respectively, with a noise limit of 0.1. These values use equally mixed targets and 256 training observations in the main comparison. Matched classical features recover this benefit, identifying the complementary representations responsible for the improvement. The controlled contribution is to isolate and quantify single-view sufficiency, complementary dependencies and representation mismatch under specified learning conditions.

On the same-input synthetic target with equal component weights, fusion reduces mean summed Brier loss from 0.386 for the classical view and 0.395 for the quantum view to 0.256. Because the target combines classical relationships with outputs of the evaluated circuit, replacing that circuit while retaining the original labels tests the importance of representation alignment; the large gain disappears under this independent-circuit comparison. Retrospective selection using validation scores chooses fusion in all 30 mixed-target cases and the corresponding single view in all 20 pure-target cases. This selection operates over complete fitted procedures, while the separated-block tests further show that the fusion rule must suit the target: concatenation succeeds on the additive target, while cross-products improve the interaction target. Together, these results identify the match between the added representation, the fusion rule and the target relationships as a central consideration in classical--quantum fusion.

Fusion improves mean macro-averaged $F_1$ over the classical baseline on Wine, three-class Fashion-MNIST and Steel Plates Faults. In the matched calibrated setting, attention increases the Wine mean from 0.963 to 0.975 and the breast-cancer mean from 0.970 to 0.972, attaining the highest breast-cancer mean among the models evaluated under that protocol. Classical controls remain competitive on Wine and breast cancer; classical models lead on Digits and Covertype; and outcomes with missing or corrupted views vary with the available inputs.

The study makes three contributions.
\begin{enumerate}
\item Controlled evidence that distinguishes single-view sufficiency from complementary target dependencies across three to five angular inputs, target mixtures and fusion rules, with matched classical features identifying the representations responsible for the diagnostic gains.
\item A complementarity framework and comparison protocol that preserve the full classical input, combine it with quantum-derived features through concatenation or attention, and assess both views through single-view comparisons, classical replacement, quantum-view masking and an independently chosen quantum transformation.
\item Validation across six single-source benchmarks and two native-view datasets that identifies when fusion improves prediction and how outcomes vary with the representation, fusion strategy and available inputs.
\end{enumerate}
Section~\ref{sec:related} introduces the quantum concepts needed to read the paper and reviews related work. Section~\ref{sec:method} defines the framework and controls. Section~\ref{sec:experiments} presents the experiments, followed by the discussion and conclusion.

\section{Background and related work}\label{sec:related}
\subsection{From classical inputs to quantum-derived features}\label{sec:quantum_background}
The quantum branch transforms a numerical input into a feature vector through encoding, a sequence of quantum operations and measurement readout. The following definitions establish the concepts needed to understand this transformation without assuming prior quantum-computing knowledge \cite{NielsenChuang2010,PQCs2019_Benedetti}.

A \emph{qubit}, or quantum bit, is a two-level system whose reference states are written $|0\rangle$ and $|1\rangle$. The vertical-bar notation denotes a state vector, which describes a pure state; a mixed state instead represents a statistical mixture of pure states. A single-qubit pure state has the form
\begin{equation}
 |\psi\rangle=\alpha|0\rangle+\beta|1\rangle,
 \qquad \alpha,\beta\in\mathbb C,\quad |\alpha|^2+|\beta|^2=1.
 \label{eq:qubit}
\end{equation}
Here $\alpha$ and $\beta$ are complex amplitudes; their squared magnitudes are the probabilities of the two outcomes when measured in this reference basis. When both amplitudes are nonzero, the state is a \emph{superposition}, whose \emph{relative phase} is the relative complex angle between the amplitudes. Subsequent operations can combine these amplitudes constructively or destructively, producing \emph{interference} that affects measurement probabilities.

\begin{figure}[htbp]
\centering\includegraphics[width=\linewidth]{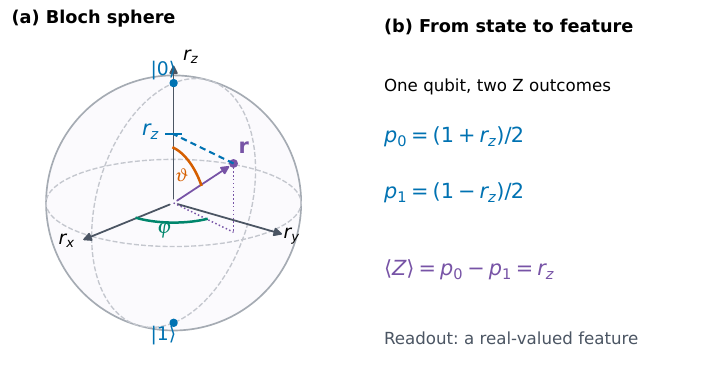}
\caption{Single-qubit geometry and readout. (a) The angles $\vartheta$ and $\varphi$ locate a pure state on the Bloch sphere; the purple arrow is its Bloch vector. The blue guide marks its vertical coordinate $r_z$. (b) This coordinate determines the probabilities of the two $Z$ outcomes and their signed mean. Mixed local states lie inside the sphere. The experiments extend this readout principle to multi-qubit local and pairwise observables.}
\label{fig:bloch}
\end{figure}

The \emph{Bloch sphere} gives a geometric description of one qubit \cite{NielsenChuang2010}. Ignoring a common phase factor, which leaves measurement probabilities unchanged, its pure state can be written
\begin{equation}
 |\psi\rangle=\cos(\vartheta/2)|0\rangle+
 e^{\mathrm i\varphi}\sin(\vartheta/2)|1\rangle,
 \quad \vartheta\in[0,\pi],\quad \varphi\in[0,2\pi).
 \label{eq:blochstate}
\end{equation}
Here $\mathrm i^2=-1$, $\vartheta$ is the polar angle measured from the north pole, and $\varphi$ is the azimuthal angle around the equator; both are in radians. These are state coordinates. The associated three-dimensional unit vector is
\begin{equation}
 \mathbf r=(r_x,r_y,r_z)
 = (\sin\vartheta\cos\varphi,\sin\vartheta\sin\varphi,\cos\vartheta).
 \label{eq:blochvector}
\end{equation}
The poles represent $|0\rangle$ and $|1\rangle$; equatorial states give equal probabilities for these outcomes. Figure~\ref{fig:bloch} connects the geometry to readout. Mixed local states lie inside the sphere; separate Bloch vectors omit multi-qubit correlations.

For $q$ qubits, the state has $2^q$ amplitudes, one for each binary string of length $q$. The symbol $\otimes$ denotes the tensor product used to combine individual states. A pure joint state is \emph{entangled} if it cannot be written as a tensor product of individual-qubit states. Entanglement describes the quantum state, distinct from statistical dependence between dataset columns.

A \emph{quantum gate} is an elementary operation on one or more qubits. The ideal gates used here are unitary, meaning that their matrices preserve the total squared magnitude of the amplitudes. A \emph{quantum circuit} is an ordered sequence of these gates. Single-qubit rotations are written
\begin{equation}
 R_P(a)=\exp(-\mathrm i aP/2),\qquad
 Y=\begin{pmatrix}0&-\mathrm i\\\mathrm i&0\end{pmatrix},\quad
 Z=\begin{pmatrix}1&0\\0&-1\end{pmatrix}.
 \label{eq:rotations}
\end{equation}
The angle $a$ is real and measured in radians, $\mathrm i^2=-1$, and $P$ is the Pauli matrix $Y$ or $Z$. The exponential is a matrix exponential. \emph{Angle encoding} assigns input-dependent values to these rotation angles. A controlled-NOT (CNOT) gate acts on two qubits by flipping the target bit when the control bit is one in the reference basis and leaving it unchanged otherwise. Acting on suitable superpositions, this operation can produce entanglement. Removing all such gates from the circuits considered here leaves a product circuit, with no operation coupling different qubits.

The gates $R_Y(a)$ and $R_Z(a)$ rotate the Bloch vector about its $y$ and $z$ axes. Starting from $|0\rangle$, $R_Y(a)$ prepares $\cos(a/2)|0\rangle+\sin(a/2)|1\rangle$, whose $Z$ expectation is $\cos a$. This illustrates how an input-dependent rotation produces a nonlinear numerical feature. The encoding angles used later are distinct from the spherical coordinates $\vartheta$ and $\varphi$.

A \emph{parameterized} or \emph{variational} circuit contains adjustable gate angles that are updated during training, whereas a fixed circuit retains its assigned angles. A \emph{layer} is a repeated gate group; \emph{depth} counts sequential stages; and \emph{ansatz} denotes the gate arrangement.

An \emph{observable} specifies a measurement and is represented by a Hermitian matrix, equal to its own conjugate transpose. For a state $|\psi\rangle$ and observable $O$, its expectation value is
\begin{equation}
 z_O=\langle\psi|O|\psi\rangle,
 \label{eq:expectation}
\end{equation}
where $\langle\psi|$ is the conjugate transpose of $|\psi\rangle$. A \emph{readout} is the numerical statistic supplied to the classical predictor: here, an expectation value. For a single-qubit $Z$ measurement, outcome $0$ is assigned value $+1$ and outcome $1$ value $-1$. If their probabilities are $p_0$ and $p_1$, then
\begin{equation}
 z_Z=p_0-p_1=r_z,\qquad
 p_0=(1+r_z)/2,\quad p_1=(1-r_z)/2.
 \label{eq:zreadout}
\end{equation}
Thus $z_Z$ is the vertical Bloch coordinate. For example, $p_0=0.8$ and $p_1=0.2$ give readout $0.6$. Each readout retains one selected statistic of the state.

For distinct qubit indices $j,k\in\{1,\ldots,q\}$, we use local $Z_j$ readouts and joint $Z_jZ_k$ readouts, with identity operations on the other qubits. The latter average the product of the two signed outcomes: $+1$ for equal bits and $-1$ for unequal bits. This joint statistic can describe dependence that separate local means omit, although a nonzero product expectation alone does not establish entanglement. Each readout lies in $[-1,1]$ and becomes one ordinary real-valued input feature.

On hardware, a \emph{shot} is one execution and measurement; averaging finitely many shots introduces sampling variation. A classical state-vector simulator stores the amplitudes and computes the expectations directly. Our main experiments use exact simulated expectations, with no simulated device noise. Supplementary Section~S8 additionally reports a separate ideal finite-shot sensitivity for the two-input diagnostic. Their purpose is to evaluate the predictive effect of the representation under these conditions.

\subsection{Sequential hybrids and retained-input fusion}
A hybrid learning system combines quantum transformations with classical processing, with sequential architectures passing the output of one stage to the input of the next. Mari et al. study transfer between classical and quantum networks and introduce dressed circuits with trainable classical transformations before and after the circuit \cite{Mari2020Transfer}. Henderson et al. use local circuit transformations as feature filters within a classical convolutional architecture \cite{Henderson2020Quanvolution}. Quantum kernel methods instead calculate similarities between encoded quantum states for a classical kernel learner \cite{Schuld2019Feature,Havlicek2019Features}. Khatun and Usman combine a pretrained classical network with a variational quantum model and report improved learning and adversarial robustness in their evaluated image-classification simulations \cite{Khatun2025Transfer}. These examples differ in the placement and role of the quantum component; classical optimization can train a hybrid even when its forward prediction follows a sequential route \cite{PQCs2019_Benedetti,Cerezo2021Variational}.

Figure~\ref{fig:routes} summarizes three representative serial arrangements and retained-input fusion. Readout-side bypass work already preserves classical features alongside the quantum outputs \cite{Zhang2025ReadoutBypass_a}. Our framework uses a multi-view formulation to compare concatenation and attention, with single-view comparisons and replacement and masking controls that assess the contributions of both views and distinguish overall fusion gains from the effect of the selected quantum transformation. The constructed-target and independent-circuit comparisons additionally examine alignment with the relationships needed for prediction. These controlled comparisons, together with the attribution controls, distinguish the study from proposing retained-input fusion or trigonometric expressivity alone.

\subsection{Multi-view learning and fusion}
Multi-view learning combines descriptions of the same observation \cite{Xu2013Survey,Baltruaitis2018MultimodalReview_a}. Native views come from separately available measurements or feature sets; generated views apply additional transformations to an existing input. Single-source multi-view methods establish that generated views can also be useful \cite{Wang2011SingleSourceMVL}. Common-subspace methods learn shared discriminative coordinates \cite{Lin2023CommonSubspace}. Multi-view sparse representations have also supported robust visual tracking \cite{Kang2019Tracking}.

Fusion can occur before prediction by appending features (early fusion), within learned representations (intermediate or latent fusion), or after separate predictors have produced class scores (late fusion) \cite{Baltruaitis2018MultimodalReview_a}. Attention pooling learns weights for a collection of feature vectors before combining them \cite{Ilse2018Pooling,Lee2019SetTransformer}. In our model, a learned summary of the classical input determines those weights. This is an application of established attention mechanisms to the classical--quantum comparison.

Trusted Multi-View Classification (TMC) learns nonnegative support, called evidence, for each class from each view. A Dirichlet distribution describes uncertainty over the class probabilities. TMC combines the views through Dempster--Shafer evidence fusion, a rule that combines class support and uncommitted uncertainty \cite{Han2021TMC}. Trusted Cross-view Completion (TCVC) learns relationships between available sources to estimate missing representations before evidence-based classification \cite{Zhou2025TCVC}. Hierarchical mutual distillation coordinates predictions from different subsets of views \cite{Yang2026HMDMV}. These approaches motivate testing an added circuit view within otherwise unchanged native-view fusion procedures.

\subsection{Representation choice and classical alternatives}
A circuit's input encoding, adjustable gates and measured observables jointly determine its feature map, meaning the transformation from input to feature vector. For the single angle-encoding layer used here, outputs have a trigonometric structure involving sine and cosine functions of the encoded coordinates \cite{Schuld2021Encoding}. Classical trigonometric features are therefore relevant comparisons. A classical surrogate is a model designed to reproduce a quantum model's input--output behaviour; existing work explains both opportunities and limitations of such alternatives \cite{Schreiber2023Surrogates,Sweke2023RFF}.

Learning from classical examples can also reduce the difficulty of predicting a quantum-generated target \cite{Huang2021Data}. We therefore evaluate complementarity relative to the specified classical representation and training procedure.

\section{Models and comparison design}\label{sec:method}
\subsection{What constitutes a useful added view?}
Let $x\in\mathbb R^d$ be an observation with $d$ real-valued input features, and let $y\in\{1,\ldots,K\}$ be its label among $K$ classes. A classical representation $c(x)\in\mathbb R^{d_C}$ and an added representation $z(x)\in\mathbb R^{d_Q}$ contain $d_C$ and $d_Q$ features, respectively. We denote the associated branches by C and Q. On native-view data, $x$ denotes the available collection of measured views. An added representation is useful when the complete procedure trained with it has lower expected test loss than the chosen reference procedure.

For a training procedure $\mathcal A$, including preprocessing, tuning and fitting, define
\begin{equation}
 \begin{aligned}
 R_n(\mathcal A)&=\mathbb E_{D_n,\omega,(x,y)}
       \bigl[\ell(\mathcal A(D_n;\omega)(x),y)\bigr],\\
 U_{Q\mid C}&=R_n(\mathcal A_C)-R_n(\mathcal A_{CQ}).
 \end{aligned}
 \label{eq:utility}
\end{equation}
Here $\ell$ is the prediction loss, $\mathcal A_C$ is the classical training procedure, $\mathcal A_{CQ}$ is the procedure using both branches, $D_n$ is a training sample of size $n$, $\omega$ represents training randomness, and $(x,y)$ is an independent test observation from the same target population. The expectation averages over all three indicated sources of variation. Positive $U_{Q\mid C}$ means that adding Q reduces expected test loss for these procedures; zero or negative values indicate no reduction or an increase. Because both representations are computed from $x$, utility concerns learnability from finite samples rather than acquisition of a new measurement. The experiments estimate this comparison for specified procedures.

\subsection{Controlled representation diagnostic}\label{sec:diagnostic_method}
Controlled regression isolates the relationships accessible through specified views to distinguish single-view sufficiency from complementary target dependencies. It compares a degree-two polynomial representation, exact two-qubit readouts, their weighted feature fusion, a nonlinear raw-input baseline and matched classical trigonometric features. The output is a real target, and squared-error loss replaces classification loss in Eq.~\eqref{eq:utility}. Training-only cross-validation selects ridge penalties, readout coefficients and fusion weights. These fixed-feature regression models establish properties of the controlled representations, while the benchmark experiments separately evaluate the nine-qubit variational classifiers and attention pooling defined below. Section~\ref{sec:sine_diagnostic} gives the construction and Supplementary Section~S8 gives the complete protocol.

\subsection{Classical input and circuit features}
In the matched neural study, the classical branch retains all input features, standardized by subtracting training means and scaling by training standard deviations. The circuit branch uses principal component analysis (PCA), a linear transformation that retains directions of greatest training-input variation, to obtain nine coordinates. PCA is fitted on the training partition. For component index $j\in\{1,\ldots,9\}$, let $t_j$ be its score and $\sigma_j$ its standard deviation computed on the training observations using the population convention (divisor equal to the number of training observations). The calibrated angle, in radians, is
\begin{equation}
 \phi_j=\frac{\pi}{2}\tanh\!\left(\frac{t_j}{\max(\sigma_j,10^{-12})}\right).
 \label{eq:angles}
\end{equation}
Here $\tanh$ is the hyperbolic tangent, bounding each calibrated angle to $(-\pi/2,\pi/2)$. The original encoding, $\phi_j=\pi\tanh(t_j)$, is evaluated separately as a sensitivity analysis. Changing between these encodings changes both scale and range.

Using the gate and state notation in Section~\ref{sec:quantum_background}, the circuit encodes the input in the product state $\bigotimes_{j=1}^{q}R_Y(\phi_j)|0\rangle$ and applies three variational layers. Each layer contains three single-qubit rotations about the $Z$, $Y$ and $Z$ axes, in that order, per qubit and a ring of controlled-NOT gates. With $q=9$, there are 81 trainable rotation parameters. The output vector contains $2q=18$ exact expectation values, comprising the local $Z_j$ measurements and the adjacent $Z_jZ_{j+1}$ measurements, with cyclic indexing $q+1\equiv1$. The vector of angles is $\phi=(\phi_1,\ldots,\phi_q)^{\mathsf T}$, and the resulting readout vector is $z(x)\in[-1,1]^{2q}$. Supplementary Section~S2 specifies the rotation order, initialization and training settings.

All circuit outputs in this study are calculated on a classical simulator. For the single input encoding used here, each readout can be expressed as a linear combination of products of $1$, $\sin\phi_j$ and $\cos\phi_j$ \cite{Schuld2021Encoding}. This constrained nonlinear structure motivates classical trigonometric controls, while allowing different feature choices to have different fitting behaviour and computational costs.

\subsection{Attention-based feature fusion}
A classical encoder maps the full input to a summary token $t_0\in\mathbb R^D$. A token is a learned feature vector, and $D$ is its width; all tokens below are treated as row vectors. Each scalar circuit readout $z_a$ is embedded as
\begin{equation}
 t_a=wz_a+b+e_a,\qquad a=1,\ldots,2q,
 \label{eq:tokens}
\end{equation}
where $w,b\in\mathbb R^D$ are shared weights and offsets, and the learned identity vector $e_a\in\mathbb R^D$ identifies observable $a$. The joint token matrix is $T=[t_0;t_1;\ldots;t_{2q}]\in\mathbb R^{(2q+1)\times D}$. An attention head is one independently parameterized weighting operation. For head $h\in\{1,\ldots,H\}$, where $H$ is the number of heads and $d_h$ their width, the summary supplies the query and the joint sequence supplies the keys and values according to
\begin{equation}
 u_h=\operatorname{softmax}\left(
 \frac{(t_0W_h^{Q})(TW_h^{K})^{\mathsf T}}{\sqrt{d_h}}
 \right)(TW_h^{V}).
 \label{eq:pool}
\end{equation}
Compatibility between the summary query and each token's key determines its attention weight, while the corresponding value supplies the features used in the pooled representation. The learned matrices $W_h^Q,W_h^K,W_h^V\in\mathbb R^{D\times d_h}$ form the query, keys and values, respectively; the superscript $Q$ in this equation means query, rather than the circuit branch. The transpose is denoted by $\mathsf T$. Softmax exponentiates the compatibility scores and normalizes them to weights summing to one, so $u_h\in\mathbb R^{d_h}$ is a weighted feature summary. The head outputs are concatenated, projected and added to the summary through a residual connection. Layer normalization rescales the summary coordinates; a feed-forward residual block then produces the final summary used for classification. The matched model uses token width $D=16$ and four heads of width $d_h=4$.

\begin{figure}[htbp]
\centering\includegraphics[width=\linewidth,height=0.76\textheight,keepaspectratio]{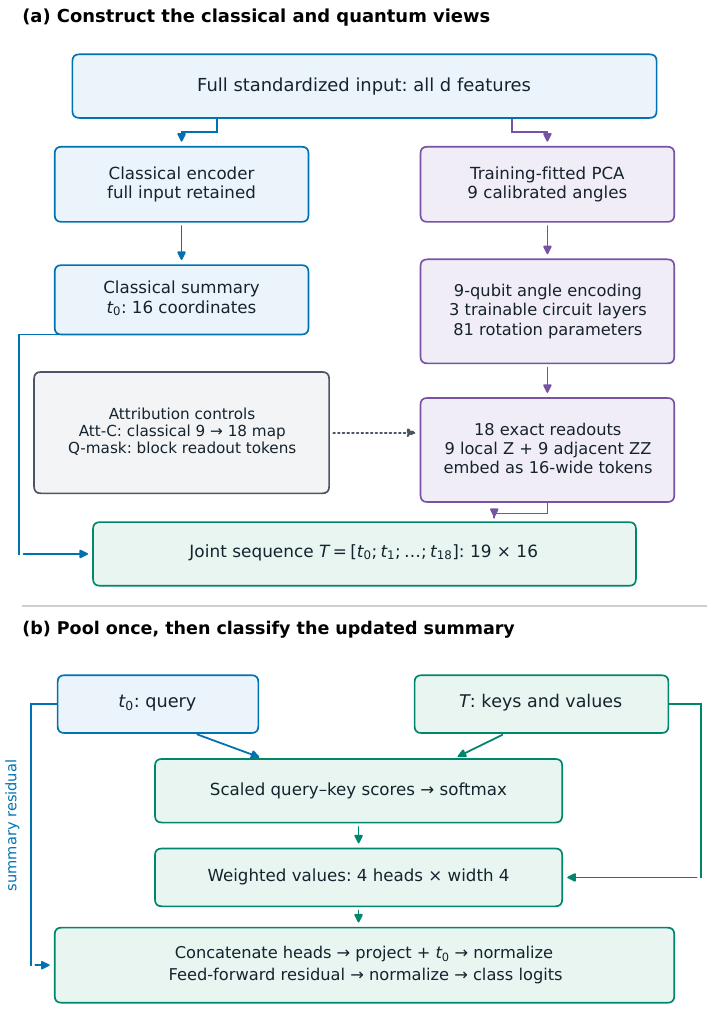}
\caption{Matched attention-fusion model. (a) Standardization preserves all $d$ classical input features; the circuit branch compresses the input to nine coordinates. Three trainable circuit layers produce 18 exact readouts. (b) The classical summary provides each attention query; all 19 tokens provide keys and values. Four head outputs update the summary through projection, residual addition, normalization and a feed-forward block. Only that summary is classified. Blue denotes the retained classical path, purple the circuit branch and teal fusion. Att-C replaces the circuit at its 18-feature interface; Q-mask blocks its tokens from the summary.}
\label{fig:framework}
\end{figure}

The classifier receives only the summary updated by this single pooling block. In this summary-query attention mechanism, the classical summary determines how the feature tokens are combined, without a subsequent block exchanging information between updated readout tokens. Figure~\ref{fig:framework} shows the model and the two main attribution controls.

Algorithm~\ref{alg:fusion} specifies training and prediction for the matched attention model while retaining the experimental separation between fitting and monitoring observations. The classifier produces real-valued class scores, or logits, which softmax converts to probabilities. Training minimizes cross entropy, the mean negative log probability assigned to the true class, using AdamW, an adaptive gradient optimizer with decoupled weight decay. The five-benchmark and fixed-circuit native-view protocols retain their own settings in Supplementary Sections~S2 and S5.

\begin{algorithm}[t]
\caption{Matched classical--quantum attention fusion}\label{alg:fusion}
\begingroup\singlespacing\small
\noindent\textbf{Input:} Fitting observations and labels, a disjoint monitoring set, an initialization seed and the chosen epoch ceiling. Each observation has $d$ features and one of $K$ labels. The calibrated setting uses nine qubits and at most 240 epochs.\par\smallskip
\begin{tabularx}{\linewidth}{@{}r@{\quad}X@{}}
1 & Fit standardization and nine-component PCA on the fitting set; retain those transformations for monitoring and prediction.\\
2 & Initialize the classical encoder, circuit angles, token embeddings, attention projections and classifier.\\
3 & \textbf{For each epoch and shuffled fitting minibatch:}\\
4 & \quad Retain the full standardized input; obtain circuit angles from the PCA scores using Eq.~\eqref{eq:angles}.\\
5 & \quad Encode the nine qubits, apply the three circuit layers and calculate the 18 local/pair expectation values.\\
6 & \quad Form the classical summary $t_0$ and readout tokens $t_a$ using Eq.~\eqref{eq:tokens}.\\
7 & \quad Pool the joint tokens with the four summary queries in Eq.~\eqref{eq:pool}; apply projection, residual connections, normalization and the feed-forward block.\\
8 & \quad Map the updated summary to $K$ logits and minimize mean cross entropy. Update all trainable components with AdamW, using learning rate and weight decay $10^{-3}$ and gradient-norm clipping at one.\\
9 & \textbf{After each epoch:} evaluate monitoring cross entropy and retain the selected checkpoint under the minimum-epoch and patience rules in Section~S2.\\
10 & \textbf{Return:} the selected network and fitted preprocessing. For a held-out observation, apply steps 4--7 and the classifier without parameter updates, then convert logits to class probabilities.\\
\end{tabularx}
\endgroup
\end{algorithm}

\subsection{Controls for the source of a gain}
The matched experiments compare five models. A multilayer perceptron (MLP), a network of successive fully connected nonlinear layers, uses the full classical input. Concat-CQ appends the 18 circuit outputs to that input before classification. Att-CQ uses the attention model in Eq.~\eqref{eq:pool}. Att-C replaces the circuit with $\tanh(A\phi+b_C)$, where $A\in\mathbb R^{18\times9}$ and $b_C\in\mathbb R^{18}$ are learned weights and offsets, while preserving the output size and pooling architecture. Q-mask retains the circuit computation but blocks all attention links through which its outputs could affect the summary.

Single-view and fused-model comparisons assess both views: the classical-only baseline tests the added quantum view, and the quantum-only baseline tests the benefit of retaining classical information. Replacement and masking further examine these contributions within the matched protocol.

Att-C has 180 branch parameters, compared with the circuit's 81, so the replacement has more than twice as many adjustable parameters at the same feature interface. This provides a learned classical alternative without restricting it to the circuit's parameter count. The MLP and concatenation hidden widths are chosen to approximate the total attention parameter count. Q-mask retains the circuit computation while isolating the predictive contribution of the classical path. Each model is fitted independently under the same data and monitoring protocol.

For native-view experiments, the classical model pools embeddings from the measured views, and the additional branch supplies 12 fixed six-qubit readouts from their concatenated inputs. A product-circuit control removes the controlled-NOT gates while retaining the remaining transformations. Because these native-view circuits are fixed and the associated attention model differs from the matched single-source model, their results are reported separately.

\subsection{Evidence fusion for native views}\label{sec:evidence_method}
TMC combines the class-evidence vectors learned from individual measured views to obtain a joint prediction \cite{Han2021TMC}. Each evidence value quantifies nonnegative support for a class and becomes a Dirichlet parameter after adding one. The resulting distribution describes uncertainty over possible class-probability vectors: its mean supplies the predicted probabilities, while the total evidence determines the model's uncommitted uncertainty. Fusion reinforces class support shared across views, with zero evidence assigned to a missing view so that its absence leaves the combined evidence unchanged. Supplementary Section~S5 gives the equations and training objective.

The comparison between TMC, which uses only the measured views, and TMC$_Q$ tests whether an additional circuit-derived representation improves evidence-based prediction. TMC$_Q$ supplies a further evidence branch with the twelve fixed circuit readouts computed from available or imputed measurements. Because it is derived from the measured views, it supplies no independent measurement; model uncertainty is assessed through predictive losses rather than assumed to be calibrated. TCVC additionally estimates missing representations from observed sources before classification; its observed-only adaptation fits preprocessing and completion targets using available fitting entries.

\subsection{Fusion-stage comparison}
The five-benchmark experiment compares six model families, comprising a full-input classical network, a circuit-only classifier and four fusion routes. Early fusion appends circuit readouts to the complete standardized classical input. Latent fusion mixes learned classical and circuit representations through a trainable scalar gate, meaning a weight constrained between zero and one. Attention fusion uses a classical summary token to pool the circuit tokens. Late fusion mixes the two branches' logits, the class scores before conversion to probabilities. The latent, attention and late routes use PCA-compressed classical inputs in this experiment; their representations and training settings therefore differ from the matched full-input controls. Section~S2 specifies the five-fold, 30-epoch protocol, and Section~S7 reports all its retained scores. These comparisons characterize the role of fusion stage within the same study.

\section{Experiments}\label{sec:experiments}
\subsection{Datasets, protocols and interpretation}
The evaluation has four parts, comprising constructed problems with known target relationships; a five-benchmark comparison of fusion stages; matched neural comparisons on Wine, Wisconsin Diagnostic Breast Cancer (WDBC) and Digits; and native-view studies on Multiple Features and a Caltech-UCSD Birds (CUB) image--text feature benchmark. All are experiments in this study. Wine contains 178 observations, 13 features and three classes; WDBC contains 569 observations, 30 features and two classes \cite{uci_wine,uci_wdbc}. The Digits experiment uses a fixed stratified subset of 900 observations from the 1,797-record scikit-learn dataset, with 64 features and ten classes \cite{digits,sklearnDigits,Pedregosa2011}.

The five-benchmark comparison also includes Steel Plates Faults (1,941 observations, 27 features and seven classes), a three-class Covertype subset with a reported cap of 5,000 observations, and a three-class Fashion-MNIST subset with a reported cap of 3,000 observations \cite{uci_steel,uci_covertype,xiao2017fashion}. These subset scores should be interpreted at their evaluated class and sample scope.

The native-view studies use all 2,000 Multiple Features observations, each described by six feature sets \cite{mfeat}. The CUB experiment uses a derived ten-class benchmark of 600 objects with aligned 1,024-dimensional image and 300-dimensional text features. It is not an evaluation on the full CUB-200-2011 image collection \cite{Wah2011CUB}. Dataset sources, subset definitions and provenance limitations are given in Supplementary Section~S1.

Preprocessing and model selection use training and validation data, not outer-test labels. The protocols reflect the different purposes of the comparisons, with synthetic experiments using independently generated datasets. The regression diagnostic reports expected-loss root mean squared error (RMSE) and repetition-level uncertainty, separately from the classification metrics. The fusion-stage experiment uses five stratified folds and a 30-epoch ceiling. Matched neural experiments use three outer folds and two initializations. Native-view attention uses three repetitions of three-fold validation, while the incomplete-view extension uses three grouped holdouts. These are separate comparisons, not one pooled benchmark ranking.

We report macro-$F_1$ (the mean of class-specific harmonic means of precision and recall), accuracy (the fraction classified correctly), area under the receiver operating characteristic curve (AUC), log loss and summed Brier loss as appropriate. AUC measures ranking over decision thresholds. Log loss penalizes low probability assigned to the true class; summed Brier loss measures squared error across all predicted class probabilities. Standard deviation is abbreviated SD. Definitions and aggregation units appear in Supplementary Section~S1. Lower loss and higher classification scores are preferred. Tables identify the unit used for their standard deviations. Bold identifies only the best mean within each dataset, target and condition. The same-input classification, ten-repetition block-target and angular--polynomial tables use retained unrounded means, with numerical ties within $10^{-12}$ retained; other tables retain ties at displayed precision. Consequently, equal rounded values need not both be bold. Bold does not indicate statistical significance. Expected Bayes loss is the expected loss of the true label-generating probabilities. These losses are reference values and are excluded from fitted-model rankings. Variation across overlapping folds is descriptive and is not treated as variation across independent datasets. The encoding and training-budget extensions are sensitivity analyses, not independently registered confirmatory studies.

\subsection{Angular relationships and an independent polynomial target}\label{sec:sine_diagnostic}
The first controlled study examines whether a selected quantum readout captures a sine--cosine relationship, whether this behaviour persists as the number of inputs increases, and whether an additional polynomial dependency creates a benefit from fusion. The initial test evaluates a two-input construction. The extension uses $d\in\{3,4,5\}$ independent angles $x_i\sim\operatorname{Unif}[-\pi,\pi]$ and an independent input $v\sim\operatorname{Unif}[-1,1]$. Here $i\in\{1,\ldots,d\}$ indexes angles, and $\operatorname{Unif}$ denotes a uniform distribution on the stated interval. The complete input therefore has four, five or six coordinates. Each pair $1\leq i<j\leq d$ has a separate two-qubit preparation
\begin{equation}
 \begin{aligned}
 |\psi_{ij,\theta}\rangle
 &= [R_Y(x_i)\otimes R_Y(x_j)]
       (\cos\theta|00\rangle+\sin\theta|11\rangle),\\
 Q_{ij,\kappa}(x)
 &=\langle Z\otimes Z\rangle
 =\cos x_i\cos x_j+\kappa\sin x_i\sin x_j,
 \end{aligned}
 \label{eq:sine_readout}
\end{equation}
where $\theta\in[0,\pi/4]$ controls the initial two-qubit state and $\kappa=\sin(2\theta)\in\{0,0.5,1\}$. Local rotations preserve the pair's entanglement. Its concurrence, a pure-state two-qubit entanglement measure ranging from zero for product states to one for maximally entangled states, is $\kappa$. This is a collection of pair circuits, not one $d$-qubit entangled state. Increasing $d$ increases the pair count from three to six to ten, while interaction order remains two.

With pair count $m=d(d-1)/2$ and fixed signs $b_{ij}\in\{-1,1\}$, define
\begin{equation}
 \begin{aligned}
 g_{d,\kappa}(x)&=\frac{2\sum_{i<j}b_{ij}Q_{ij,\kappa}(x)}{\sqrt{m(1+\kappa^2)}},\\
 p(v)&=\frac{\sqrt3v+\sqrt5(3v^2-1)/2}{\sqrt2},\\
 y&=\sqrt w\,g_{d,\kappa}(x)+\sqrt{1-w}\,p(v)+\epsilon,
 \end{aligned}
 \label{eq:sine_target}
\end{equation}
Here $g_{d,\kappa}$ is the angular signal and $p$ is the polynomial signal. The independent noise $\epsilon\sim\mathcal N(0,0.1^2)$ is Gaussian with mean zero and standard deviation 0.1; $w\in\{0,0.5,1\}$ is the angular signal's share of target variance. Both components have mean zero and variance one. Pure polynomial, equally mixed and pure angular targets thus have the same signal variance. The polynomial input is retained in every condition, including those where its target coefficient is zero.

C uses all degree-one and degree-two polynomial features of the complete input. Q uses the pair readouts and selects its sine coefficient using training data; this chosen view omits $v$. C+Q selects a weighted combination of the two feature kernels. A classical union computes the same trigonometric features without quantum simulation and combines them with C using equal block weights. A feature kernel measures similarity by an inner product of feature vectors. All feature models use ridge regression, which fits squared error with a penalty on squared coefficient size; a Gaussian radial basis function (RBF) kernel provides a nonlinear full-input comparison. Twenty independent repetitions use 128 or 256 training observations and 2,048 test inputs. Three-fold training validation selects all settings. Expected-loss root mean squared error (RMSE) integrates the known observation-noise variance, giving a lower limit of 0.1. The study is exploratory; full grids, uncertainty estimates and reversals across sine coefficients are retained in Section~S8.

\begin{table}[t]
\centering\begingroup\singlespacing\small
\caption{Controlled regression with 256 training observations, full sine coefficient $\kappa=1$ and 20 independent repetitions. Mean expected-loss RMSE; lower is better. C is polynomial, Q is the selected pair-readout view, and C+Q is feature-kernel fusion. Union is the matched classical combination; RBF uses all raw inputs. The noise limit is 0.1. These are the diagnostic models of Section~\ref{sec:diagnostic_method}. Repetition SDs and 95\% intervals are in Section~S8.}
\label{tab:sine_main}
\setlength{\tabcolsep}{5pt}
\renewcommand{\arraystretch}{1.12}
\begin{tabular}{@{}lrrrrrr@{}}\toprule
Target&$d$&C&Q&C+Q&Union&RBF\\\midrule
Polynomial&3&\textbf{0.1030}&1.0128&0.1033&0.1038&0.1332\\
&4&\textbf{0.1043}&1.0132&0.1048&0.1056&0.1544\\
&5&\textbf{0.1059}&1.0141&\textbf{0.1059}&0.1081&0.1815\\\midrule
Angular&3&0.9330&\textbf{0.1008}&0.1011&0.1036&0.3604\\
&4&0.9378&\textbf{0.1015}&\textbf{0.1015}&0.1055&0.5827\\
&5&0.9465&\textbf{0.1021}&\textbf{0.1021}&0.1078&0.7752\\\midrule
Mixed&3&0.6651&0.7233&0.1037&\textbf{0.1037}&0.3373\\
&4&0.6779&0.7241&0.1056&\textbf{0.1056}&0.4564\\
&5&0.6866&0.7293&0.1080&\textbf{0.1079}&0.6584\\
\bottomrule\end{tabular}\endgroup
\end{table}

The reproduced two-input test confirms that the aligned readout captures the sine contribution missed by the cosine-only readout; matched classical features reproduce its predictions (Section~S8). In the dimension extension, C suffices for the pure polynomial target, and Q suffices for the aligned angular target (Table~\ref{tab:sine_main}). When both components determine the target, fusion restores near-noise-limit prediction across all three dimensions: mean RMSE is 0.1037, 0.1056 and 0.1080 at three, four and five angular inputs for equally mixed targets, the full sine coefficient and 256 training observations. The classical union closely matches fusion. The result therefore identifies complementary target relationships accessible through the selected feature sets and regression heads.

\subsection{Changing the target while keeping the inputs fixed}\label{sec:sameinput}
The same-input constructed study gives every model the same nine input coordinates and changes only the label rule within each repetition. C contains the angles and their sine and cosine values, giving 27 features. Q contains 18 fixed circuit readouts. C+Q concatenates these features and fits an ordinary logistic classifier.

Let $\bar c(x)$ be the standardized score $x_1/\pi+0.7\sin x_2-0.4\cos x_3$, and let $u(x)$ be the standardized ninth local-$Z$ readout of the generating circuit. For $w\in\{0,0.25,0.5,0.75,1\}$, labels are drawn with probability
\begin{equation}
 \eta_w(x)=\sigma\!\left(3\frac{s_w(x)-\mu_w}{\sigma_w}\right),
 \qquad s_w(x)=\sqrt{1-w}\,\bar c(x)+\sqrt w\,u(x).
 \label{eq:sameinput}
\end{equation}
Here $\sigma(t)=1/(1+\exp(-t))$ is the logistic function for a real-valued argument $t$. The weight $w$ controls the contribution of the circuit-derived score to the label rule; $\mu_w$ and $\sigma_w>0$ are the mean and standard deviation of $s_w$ on 6,000 independent unlabeled calibration observations. Thus changing $w$ changes the prediction target, not the physical nature of the input data. An additional condition uses independent fair random labels. Ten data/circuit repetitions each contain 600 development and 1,200 test observations. Three-fold inner validation selects logistic regularization before refitting on the development data.

An independent-circuit control, $Q_{\rm ind}$, uses the same architecture and readout count, but a different fixed draw of rotation parameters. Two compact classical controls use 18 random trigonometric products (RP) or 18 integer-frequency Fourier features (RF). C+T appends first-order sine/cosine features already present in C; it changes regularization but does not enlarge the feature span. Supplementary Section~S4 gives every target condition, selection rule and seed.

\begin{table}[t]
\centering\begingroup\singlespacing\small
\caption{Same-input constructed target with $w=0.5$. Summed Brier loss over ten data/circuit repetitions. All fitted procedures use logistic classifiers; lower is better. The target includes a readout of Q, while $Q_{\rm ind}$ is an independent circuit draw. RP and RF are classical feature alternatives with the same added-feature count.}
\label{tab:sameinput}
\setlength{\tabcolsep}{12pt}
\begin{tabular}{@{}lrr@{}}\toprule
Procedure&Mean&SD\\\midrule
C&0.386&0.018\\
Q&0.395&0.014\\
C+Q&\textbf{0.256}&0.018\\
$Q_{\rm ind}$&0.503&0.004\\
C+$Q_{\rm ind}$&0.397&0.016\\
C+T&0.387&0.019\\
C+RP&0.399&0.019\\
C+RF&0.397&0.018\\\bottomrule
\end{tabular}\endgroup
\end{table}

At the equal mixture, C+Q has mean loss 0.256, compared with 0.386 for C, 0.395 for Q and 0.397 for C+$Q_{\rm ind}$ (Table~\ref{tab:sameinput}). Relative reductions calculated from the retained unrounded means are 33.7\% against C and 35.1\% against Q. The large gain disappears with the independent circuit draw, and the compact classical controls also fail to recover the generating readout under this feature budget. These comparisons support alignment between the chosen representation and the label-generating relationship as an explanation for the gain. On the pure-classical target, adding Q increases mean loss from 0.237 to 0.245; on the pure-circuit target, adding C increases it from 0.251 to 0.263. Random-label losses remain close to 0.5. The construction therefore distinguishes complementary target relationships from cases in which a single representation is sufficient and fusion provides no benefit (Supplementary Section~S4).

\paragraph{Validation-based model selection.} A retrospective comparison of the stored validation scores selects C+Q in all 30 mixed-target cases and the corresponding individual view in all 20 pure-target cases. With the independent circuit, it selects C in 29 of the 30 mixed-target cases. These are choices between complete logistic procedures, not fitted weights within a fused model. They describe this constructed study and do not validate a prospective rule for unseen datasets. All random-label and non-favourable conditions are retained in the supplement.

\subsection{Controlled targets with two feature blocks}\label{sec:controlled}
The block-target study asks whether the fusion rule can use a known relationship between two representations. Inputs have nine independent coordinates sampled uniformly from $[-\pi,\pi]$. The classical view C contains 18 low-order features of coordinates 1--3. The circuit view Q contains 18 readouts generated from coordinates 4--9. The circuit has nine qubits, encodes $(0,0,0,x_4,\ldots,x_9)$ and measures nine local $Z_j$ and nine adjacent $Z_jZ_{j+1}$ expectations with cyclic indexing. Standardized scores $a$ and $b$ are functions of these respective blocks. Binary labels are sampled with probability $\sigma(3s)$, where $\sigma(v)=1/(1+\exp(-v))$ is the logistic function for a real-valued argument $v$ and the target score is
\begin{equation}
 \begin{aligned}
 s_{\rm additive}&=\frac{a+b}{\sqrt2},\qquad s_{\rm interaction}=ab,\\
 s_{\rm Fourier}&=2\sin x_4\cos x_5.
 \end{aligned}
 \label{eq:targets}
\end{equation}
The additive head concatenates the feature blocks. The interaction head also includes their pairwise cross-products. C+replacement substitutes 18 classical trigonometric features for Q. A full-input classical trigonometric control is included separately. All displayed fitted models use logistic regression with validation-selected regularization. Each of three generated datasets contains 600 fitting, 300 validation and 1,200 test observations. The exact features, generating circuit and seeds are given in Supplementary Section~S4.

\begin{table}[htbp]
\centering\begingroup\singlespacing\small
\caption{Constructed block-target test log loss: mean (SD) over three independent generated datasets. Lower is better. The final row is the conditional expected Bayes loss, not a fitted model.}
\label{tab:controlled}
\setlength{\tabcolsep}{6pt}
\renewcommand{\arraystretch}{1.12}
\begin{tabular}{@{}lccc@{}}\toprule
Model & Additive & Interaction & Fourier\\\midrule
C & 0.561 (0.004) & 0.695 (0.003) & 0.694 (0.001)\\
Q & 0.585 (0.001) & 0.694 (0.001) & 0.693 (0.001)\\
C+Q, additive & \textbf{0.402} (0.010) & 0.696 (0.004) & 0.693 (0.001)\\
C+Q, interaction & 0.436 (0.005) & \textbf{0.666} (0.007) & 0.693 (0.001)\\
C+replacement & 0.570 (0.006) & 0.694 (0.001) & 0.580 (0.026)\\
Trig., all inputs & 0.563 (0.003) & 0.699 (0.014) & \textbf{0.504} (0.009)\\\midrule
Expected Bayes loss & 0.355 (0.003) & 0.445 (0.003) & 0.353 (0.008)\\\bottomrule
\end{tabular}
\endgroup
\end{table}

On the additive target, C+Q reduces mean log loss from 0.561 for C and 0.585 for Q to 0.402 (Table~\ref{tab:controlled}). The interaction head is needed to improve on the individual views for the interaction target, but its loss of 0.666 remains well above the expected Bayes loss of 0.445. On the Fourier target, the displayed full-input classical control reaches 0.504, compared with 0.693 for the circuit combinations. Thus, the useful representation and the useful fusion rule both depend on the target.

A ten-repetition extension retains the original three block-target datasets and adds seven data seeds and compact classical controls. At a fitting size of 600, additive C+Q has loss $0.396\pm0.010$, compared with $0.573\pm0.012$ and $0.575\pm0.013$ for equally sized random-product and random-Fourier additions. Here the teacher, meaning the label-generating rule, deliberately uses the evaluated circuit representation. The complete six-target results at both fitting sizes are retained in Supplementary Section~S4.5, including the pure-view and random-label conditions.

These positive controls demonstrate how the specified feature maps and fusion heads exploit a known complementary relationship within the conditions established by the target construction. The additive and interaction labels contain a readout of the same fixed circuit used by Q, and C and Q receive different coordinate blocks. The comparison therefore establishes predictive utility under those conditions. The same-input experiment in Section~\ref{sec:sameinput} examines representation alignment while holding available input coordinates constant.

A sample-size sensitivity keeps the interaction target and reuses its validation and test observations. Increasing the fitting set to 1,200 and 2,400 reduces the interaction-head loss to $0.609\pm0.021$ and $0.539\pm0.016$, respectively. This is consistent with a finite-sample estimation difficulty, but the reused observations prevent treating the follow-up as independent confirmation.

\subsection{Matched neural comparisons}\label{sec:neural}
The neural study compares the five models in Section~\ref{sec:method} using the same full classical inputs, nine circuit-input coordinates, outer partitions and training budgets. A monitoring subset inside each outer-training fold selects the checkpoint. Scaling and PCA are fitted without that monitoring subset or the outer-test fold. The calibrated protocol allows up to 240 epochs. Supplementary Sections~S2 and S3 give the complete training details and the original-encoding sensitivities.

\begin{table}[htbp]
\centering\begingroup\singlespacing\small
\caption{Matched neural macro-$F_1$ under the calibrated 240-epoch protocol: mean (fold SD). Each mean averages two initialization-specific scores after pooling three outer folds. SD describes the six fold scores, not the standard error of the mean. Higher is better.}
\label{tab:neural}
\setlength{\tabcolsep}{6pt}
\renewcommand{\arraystretch}{1.12}
\begin{tabular}{@{}lccc@{}}\toprule
Method & Wine & WDBC & Digits\\\midrule
MLP & 0.963 (0.031) & 0.970 (0.007) & \textbf{0.953} (0.017)\\
Concat-CQ & 0.966 (0.029) & 0.969 (0.008) & 0.944 (0.022)\\
Att-CQ & 0.975 (0.021) & \textbf{0.972} (0.009) & 0.950 (0.010)\\
Att-C & 0.977 (0.023) & 0.971 (0.007) & 0.949 (0.016)\\
Q-mask & \textbf{0.978} (0.018) & 0.971 (0.007) & 0.948 (0.014)\\\bottomrule
\end{tabular}
\endgroup
\end{table}

The matched calibrated comparison shows gains over the MLP on Wine and WDBC (Table~\ref{tab:neural}). Att-CQ raises mean macro-$F_1$ from 0.963 to 0.975 on Wine and from 0.970 to 0.972 on WDBC. On WDBC, Att-CQ has the highest displayed mean among all five models, including the two attribution controls. The corresponding increases are 0.012 and 0.002 in $F_1$. On Wine, Att-C and Q-mask reach 0.977 and 0.978, showing that the classical paths can also achieve the improvement. On WDBC, their means of 0.971 are close to Att-CQ. On Digits, the MLP leads at 0.953, compared with 0.950 for Att-CQ. The baseline gains and the control comparisons therefore answer distinct questions.

Across the three datasets and three encoding/budget settings, all nine recorded conditional 95\% intervals for the Att-CQ--Att-C difference include zero, leaving a consistent improvement over the classical replacement unresolved. These intervals condition on fitted predictions, exclude variation from refitting or tuning, and support neither an equivalence claim nor family-wide rejection decisions (Supplementary Section~S2).

The encoding sensitivity also matters. Under the original angles, concatenation gives Wine macro-$F_1$ of 0.944 at the 80-epoch ceiling and 0.952 at 240 epochs; the calibrated 240-epoch result is 0.966. On WDBC, original-angle concatenation attains 0.975 at 80 epochs and 0.974 at 240 epochs, compared with 0.970 for the matched MLP at both original-encoding budgets (O80 and O240 in Supplementary Section~S3); calibrated attention attains 0.972. Calibration therefore benefits some combinations but is not uniformly preferable. The best observed fusion rule depends on dataset and encoding, while the classical attention controls remain competitive. The complete sensitivity table retains all settings rather than selecting the most favourable one for each method.

\subsection{Native views under source degradation}\label{sec:native}
Multiple Features provides six descriptions of each digit, comprising Fourier coefficients, profile correlations, Karhunen--Lo\`eve coefficients, pixel averages, Zernike moments and morphological measurements. Attention C combines learned 32-dimensional embeddings of these sources. Attention C+Q adds 12 fixed six-qubit readouts computed from their concatenated inputs. Attention C+Qp uses the same branch without controlled-NOT gates. The models are trained on complete data and evaluated after removing or corrupting one source at a time. Missing inputs receive training-mean imputation, absent native tokens are masked, and circuit features are recomputed from the available inputs.

\begin{table}[t]
\centering\begingroup\singlespacing\small
\caption{Native-view attention on all 2,000 Multiple Features objects. Mean and SD are calculated across three repetitions, each pooling three outer folds. Missing and noisy Brier losses first average the six source-specific conditions within a repetition. All losses are summed Brier; lower is better. TMC uses its extended training sensitivity.}
\label{tab:native_attention}
\setlength{\tabcolsep}{3.2pt}
\begin{tabular}{@{}lrrrrrr@{}}\toprule
&\multicolumn{2}{c}{Clean}&\multicolumn{2}{c}{Missing}&\multicolumn{2}{c}{Noisy}\\
\cmidrule(lr){2-3}\cmidrule(lr){4-5}\cmidrule(lr){6-7}
Method&Mean&SD&Mean&SD&Mean&SD\\\midrule
TMC (extended)&\textbf{0.037}&0.001&\textbf{0.042}&0.002&\textbf{0.041}&0.001\\
Attention C&0.046&0.002&0.133&0.012&0.051&0.002\\
Attention C+Q&0.049&0.005&0.131&0.023&0.054&0.005\\
Attention C+Qp&0.047&0.006&0.134&0.017&0.052&0.006\\\bottomrule
\end{tabular}\endgroup
\end{table}

The native-view comparison contains both improvements and increases in error (Table~\ref{tab:native_attention}). Adding Q reduces the mean over missing-view conditions from 0.133 to 0.131, while increasing clean loss from 0.046 to 0.049 and noisy loss from 0.051 to 0.054. Mean loss increases in five of the six individual source-removal conditions. Removing the profile view is the exception, with paired Brier reduction $0.061\pm0.069$; the large repetition variation limits that observation. The mean over all six missing-view conditions is slightly lower with Q, illustrating why an average should be read alongside the source-specific results in Supplementary Section~S6.

The product-circuit variant has mean losses close to those of the entangling circuit relative to the reported repetition SDs, providing no clear predictive advantage for including the controlled-NOT gates under these conditions.

TMC combines class evidence from the native views as defined in Section~\ref{sec:evidence_method} \cite{Han2021TMC}. Its extended training sensitivity reaches clean Brier loss 0.037 and lower mean missing/noisy losses than the displayed attention models. Six of its nine candidate-selection runs reach the 3,000-epoch ceiling before satisfying the patience rule, although their selected checkpoints occur earlier. These values should not be presented as a fully converged optimum for TMC.

\subsection{An additional view in evidence-based fusion}\label{sec:tmc}
A separate extension compares TMC with TMC$_Q$, the additional circuit-evidence branch defined in Section~\ref{sec:evidence_method}. It uses three duplicate-grouped holdouts, keeping observations identified as duplicates in the same partition, with approximately 60\% fitting, 20\% validation and 20\% test observations. Selected models are refitted on the development data. Multiple Features is fitted with complete inputs and tested with individual sources missing. On CUB, one source is missing in half of the objects during fitting; complete inputs are restored only for the complete-input test condition.

TCVC provides a further incomplete-view comparison on CUB \cite{Zhou2025TCVC}. We use an observed-only adaptation of the released networks. Scaling and completion targets use observed fitting entries. The adaptation and its training limits are specified in Supplementary Section~S5.

\begin{table}[htbp]
\centering\begingroup\singlespacing\small
\caption{Evidence-fusion extension: mean (SD) over three grouped holdouts. The Multiple Features missing condition averages six source removals. TCVC$^\dagger$ is the observed-only adaptation. Lower summed Brier loss and higher accuracy are preferred.}
\label{tab:native_extension}
\setlength{\tabcolsep}{6pt}
\renewcommand{\arraystretch}{1.12}
\begin{tabular}{@{}llcc@{}}\toprule
Input & Method & Brier loss & Accuracy\\\midrule
\multicolumn{4}{@{}l}{Multiple Features}\\\addlinespace[2pt]
Complete & TMC & \textbf{0.032} (0.006) & 0.974 (0.008)\\
 & TMC$_Q$ & \textbf{0.032} (0.006) & \textbf{0.976} (0.007)\\
Missing & TMC & \textbf{0.036} (0.007) & 0.975 (0.007)\\
 & TMC$_Q$ & \textbf{0.036} (0.007) & \textbf{0.976} (0.006)\\\midrule
\multicolumn{4}{@{}l}{CUB image--text benchmark}\\\addlinespace[2pt]
Complete & TMC & 0.168 (0.035) & 0.892 (0.036)\\
 & TMC$_Q$ & 0.164 (0.033) & 0.892 (0.030)\\
 & TCVC$^\dagger$ & \textbf{0.123} (0.019) & \textbf{0.919} (0.017)\\
Missing & TMC & 0.274 (0.018) & 0.828 (0.034)\\
 & TMC$_Q$ & 0.272 (0.019) & 0.831 (0.017)\\
 & TCVC$^\dagger$ & \textbf{0.193} (0.033) & \textbf{0.889} (0.013)\\\bottomrule
\end{tabular}
\endgroup
\end{table}

The added circuit produces small descriptive gains in the grouped-holdout comparison (Table~\ref{tab:native_extension}). On CUB, complete-input Brier loss decreases from 0.168 to 0.164 and missing-input loss from 0.274 to 0.272; missing-input accuracy rises from 0.828 to 0.831. On Multiple Features, Brier losses are equal at displayed precision, while accuracy increases from 0.974 to 0.976 with complete inputs and from 0.975 to 0.976 with missing inputs. These changes are small relative to the reported holdout variation. On incomplete CUB, TCVC reaches Brier loss 0.193, compared with 0.274 for TMC and 0.272 for TMC$_Q$. All 48 TMC candidate fits satisfy the validation plateau rule before the 12,000-epoch ceiling; two of the three selected TCVC runs reach their 1,000-epoch ceiling. The grouped-holdout results are distinct from the repeated-fold experiment in Table~\ref{tab:native_attention} and should not be used to infer a training-budget effect across tables.

\subsection{Fusion stages across five benchmarks}\label{sec:five}
The five-benchmark experiment evaluates classical-only and circuit-only prediction, latent mixing, attention fusion, early concatenation and late score fusion under five-fold validation with a 30-epoch ceiling. Table~\ref{tab:five} reports every retained macro-$F_1$ result. It includes the same-study Fashion-MNIST and Steel Plates Faults experiments alongside Wine, WDBC and Covertype. Architecture and preprocessing differences from the matched study are specified in Section~\ref{sec:method} and Supplementary Section~S2; the two tables are not pooled into one ranking.

\begin{table}[t]
\centering\begingroup\singlespacing\footnotesize
\caption{Five-benchmark fusion-stage experiment. Macro-$F_1$ mean (fold SD) over five folds. Fashion and Steel denote Fashion-MNIST and Steel Plates Faults. Covertype and Fashion use the three-class subsets specified in the text.}\label{tab:five}
\setlength{\tabcolsep}{2pt}
\begin{tabular}{@{}lccccc@{}}\toprule
Method&Wine&WDBC&Covertype&Fashion&Steel\\\midrule
Classical&0.943 (0.079)&\textbf{0.966} (0.019)&\textbf{0.811} (0.009)&0.969 (0.007)&0.768 (0.015)\\
Quantum&0.205 (0.017)&0.405 (0.101)&0.335 (0.036)&0.491 (0.032)&0.381 (0.038)\\
CQ latent&0.844 (0.047)&0.950 (0.027)&0.771 (0.013)&0.963 (0.012)&0.640 (0.034)\\
CQ attention&\textbf{0.967} (0.036)&\textbf{0.966} (0.016)&0.780 (0.028)&0.963 (0.012)&\textbf{0.771} (0.012)\\
CQ early&0.928 (0.033)&0.960 (0.024)&0.806 (0.018)&\textbf{0.971} (0.008)&0.738 (0.025)\\
CQ late&0.852 (0.077)&0.954 (0.027)&0.794 (0.020)&0.964 (0.007)&0.717 (0.047)\\\bottomrule
\end{tabular}\endgroup
\end{table}

On Fashion-MNIST, early fusion achieves mean macro-$F_1$ $0.971\pm0.008$, compared with $0.969\pm0.007$ for the classical baseline. On Steel Plates Faults, attention fusion achieves $0.771\pm0.012$, compared with $0.768\pm0.015$. These are improvements of 0.002 and 0.003, respectively, at displayed precision. Attention also raises Wine macro-$F_1$ from $0.943\pm0.079$ to $0.967\pm0.036$. On WDBC, classical and attention scores are both 0.966 at displayed precision. On Covertype, the classical baseline leads at $0.811\pm0.009$; early fusion is the strongest fusion route at $0.806\pm0.018$. The circuit-only model has lower mean macro-$F_1$ than the classical baseline on every dataset.

The preferred fusion route also depends on the evaluation metric. Attention achieves the highest mean AUC on Wine (0.999), WDBC (0.995) and Fashion-MNIST (0.995). Fashion-MNIST's highest $F_1$ instead comes from early fusion. On Steel Plates Faults, attention improves $F_1$ but has AUC 0.942, compared with 0.948 for the classical baseline. Thus, the leading method depends on both dataset and metric. Supplementary Section~S7 retains all AUC means and SDs, including the Covertype tie at 0.914.

\begin{figure}[t]
\centering\includegraphics[width=\linewidth]{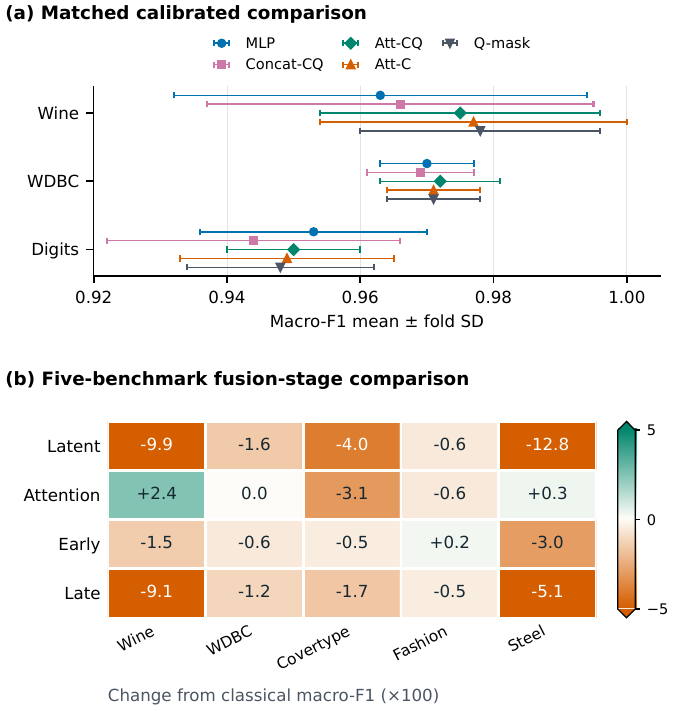}
\caption{Balanced visual summary of the retained benchmark results. (a) Matched calibrated macro-$F_1$ means and descriptive fold SDs from Table~\ref{tab:neural}; the bars are not confidence intervals. (b) Change from the classical mean for every fusion route in Table~\ref{tab:five}, multiplied by 100. Positive values favour fusion. The colour range is clipped at $\pm5$, while annotations show the actual differences of the displayed means. Fashion and Steel denote the evaluated Fashion-MNIST subset and Steel Plates Faults. Fold SDs for this panel remain in Table~\ref{tab:five}. The panels use different protocols and support separate comparisons.}
\label{fig:results}
\end{figure}

Figure~\ref{fig:results} presents the matched and five-benchmark results as separate panels. The matched panel displays every compared model and its fold SD. The fusion-stage panel shows all four fusion routes relative to the classical baseline, including the Fashion-MNIST and Steel gains and the reductions on Covertype. This display makes the dependence on dataset and fusion rule visible without combining the two protocols.

These results establish the mean improvements within the five-fold experiment. The matched experiment provides additional attribution controls under its own architecture and training protocol.

\FloatBarrier

\section{Discussion}\label{sec:discussion}
\subsection{Complementarity, representation and fusion}
The experiments show that classical--quantum fusion can improve prediction when the available representations and the fusion rule together capture relationships relevant to the target.

The angular--polynomial diagnostic establishes this relationship under controlled conditions: an aligned sine--cosine readout represents the angular component, while the polynomial view represents the independent polynomial component. Combining these representations supports accurate prediction when both components determine the target, across three, four and five angular inputs. Because the chosen quantum encoding omits the polynomial coordinate, its mixed-target error reflects a deliberately missing dependency. Classical features containing the same trigonometric relationships recover the improvement, identifying the complementary feature content responsible for the diagnostic gains. Increasing the number of inputs introduces more pair relationships without making this construction classically intractable. By specifying the target relationships and available feature sets, the experiment isolates and quantifies sufficiency, complementarity and representation mismatch under the evaluated fitting budgets.

The same-input constructed problem tests the same principle in classification. At equal target weighting, C+Q reduces mean summed Brier loss by 33.7\% relative to C and 35.1\% relative to Q. The separated-block study further distinguishes the role of representation from that of fusion: concatenation succeeds on the additive target, cross-products improve the interaction target, and a classical trigonometric representation performs better on the Fourier target. These differences show that the predictive benefit depends on both the relationships represented by the views and the capacity of the fusion rule to combine them.

Representation alignment is directly controlled in the constructed experiments because the label rule contains a readout of the circuit supplied to the classifier. Replacing that circuit with an independent draw of the same architecture removes the large gain, demonstrating the importance of the particular feature map to the observed benefit. The pure-view targets also show why fusion need not help when one representation already supports the target adequately. The remaining interaction-loss gap and its reduction with larger fitting samples illustrate that a representable relationship may still be difficult to estimate from limited data.

Retrospective model selection using the stored validation scores also distinguishes the target regimes: it selects fusion in all 30 mixed-target cases and the corresponding single view in all 20 pure-target cases. By selecting among complete fitted models, this comparison connects the controlled complementarity evidence to the choice of prediction procedure; prospective evaluation on unseen datasets is needed to assess how well the observed selection behaviour transfers.

The benchmark gains extend the empirical motivation beyond these constructed examples. In the five-benchmark experiment, the strongest fusion routes improve macro-$F_1$ over the classical baseline on Wine, Fashion-MNIST and Steel Plates Faults. In the matched calibrated comparison, attention fusion improves the MLP mean on Wine and WDBC and attains the highest WDBC mean among the models evaluated under that protocol. The dataset-specific outcomes support evaluating complementary feature fusion while showing that its practical value depends on fusion stage, encoding and evaluation metric. Fashion-MNIST favours early fusion for $F_1$ and attention for AUC, while Steel Plates Faults favours attention for $F_1$ and the classical baseline for AUC.

\subsection{What cases without improvement mean}\label{sec:no_gain}
The benefit of an added circuit varies across datasets and evaluation conditions, with the classical baseline leading on Digits and Covertype. On Multiple Features, the circuit view slightly reduces average missing-view loss but increases clean and noisy loss. In the grouped native-view experiment, several mean scores improve with Q, while other scores are equal at displayed precision. These cases establish the observed utility of the evaluated representation under each protocol.

One possible explanation is insufficient useful complementarity, arising when the selected circuit features are redundant with the classical representation or poorly aligned with the relationships required for prediction. Other possibilities include information lost during input compression, a restricted set of measured observables, limited training data, or optimization that fails to learn useful parameters. The encoding and sample-size sensitivities indicate that these design choices can affect prediction, although the experiments do not isolate a single cause for every unchanged or lower score.

These experiments assess the practical complementarity of the chosen representations; physical quantum properties of a data source are a separate question requiring measurements of the relevant quantum observables.

\subsection{Assessing both views' contributions}
A higher fused-model mean than the classical baseline documents a numerical improvement for the complete prediction procedure, while single-view comparisons assess the contribution of each representation. Replacement and masking refine this assessment by examining whether circuit features account for a gain beyond the retained classical path and a learned classical alternative. On Wine, Att-C and Q-mask have higher displayed means than Att-CQ; on WDBC, Att-CQ leads by 0.001 over these controls. The mean improvements remain part of the results, while a consistent incremental benefit over the classical replacement is unresolved by these comparisons.

Att-C uses the same feature count and fusion interface as Att-CQ but has more branch parameters. Q-mask preserves the classical pathway while preventing circuit outputs from contributing. These controls evaluate specific classical alternatives within the matched protocol, rather than establishing superiority over the full range of classical learners. The five-benchmark fusion-stage experiment uses different architectures, preprocessing and training budgets, so its Fashion-MNIST and Steel Plates Faults gains cannot be assigned the matched experiment's attribution outcome.

For native views, recomputing the circuit from available or imputed measurements transforms the observed information without supplying the independently missing source. The source-specific results show why a modest average gain should be accompanied by the individual conditions. On Multiple Features, profile-view removal favours Q on average, while the other five removals do not. The TMC extension similarly supports condition-specific mean improvements rather than a uniform robustness claim.

\subsection{Limitations and further evaluation}
All circuits are small and classically simulated, with the main-paper scores evaluating exact circuit-derived features without device noise or finite-shot variation. The separate two-input sensitivity in Section~S8 uses ideal sampled measurements, so neither protocol establishes a hardware speed-up or identifies entanglement as the cause of a predictive gain. The native product-circuit control compares gate arrangements, but is not a direct measurement of the role of entanglement during learning.

The matched experiment covers three datasets and two initializations. The five-benchmark table reports fold aggregates; original checkpoints and exact Covertype/Fashion-MNIST row identifiers are not available in the retained records. The CUB feature benchmark lacks raw-image identifiers and a documented extraction history, limiting checks of source alignment and possible feature-extractor pretraining overlap. These record limitations constrain reproduction of the exact fitted models and evaluated subsets. The accompanying code reproduces the angular--polynomial diagnostic in Section~S8; it does not reproduce the Section~S4 classification studies, their retrospective procedure-selection counts or the neural benchmark experiments, whose original training implementations have not been recovered.

The constructed targets deliberately use specified classical and circuit features, and the regression diagnostic combines target-aligned features with a larger selection grid for fusion. Its fixed pair circuits and ridge head differ from the trainable nine-qubit and attention models, whose performance on the new dimension and polynomial-mixture targets remains untested. Applying those trainable models to the same controlled targets, with matched classical alternatives and selection budgets, is the direct next test of whether the diagnostic findings transfer to the benchmark architectures. The ten-repetition classification extensions include outcomes already known when the extensions were designed and are treated as sensitivity analyses. Some native-view candidate fits reach their epoch ceilings, so the comparison does not establish a fully optimized baseline. The numerical differences reported here should be distinguished from statistical significance and practical benefit, especially for high baseline scores.

A stronger prospective evaluation would choose datasets and model-search budgets before examining test outcomes, compare multiple circuit encodings and readout sets, and include classical feature alternatives with matched training effort. An independently chosen circuit is already tested here; training such circuits under the same selection budget would further separate fixed representation mismatch from optimization effects. Evaluating feature generation, fitting time and finite-shot performance alongside predictive metrics would establish when an observed improvement justifies the additional resources. These evaluations would extend the present comparison from predictive utility to the conditions required for practical deployment.

\section{Conclusion}\label{sec:conclusion}
The controlled experiments show when classical--quantum fusion improves prediction: the selected views capture complementary target relationships, and the fusion rule can combine them. Across three, four and five angular inputs, the regression diagnostic distinguishes targets adequately captured by one selected view from mixed angular--polynomial targets that benefit from combining views. On equally mixed targets with the full sine coefficient and 256 training observations, fusion achieves mean root mean squared errors of 0.1037, 0.1056 and 0.1080 against a noise limit of 0.1. Matched classical features reproduce this improvement, identifying representation complementarity as the source of the diagnostic gains.

The same-input classification test supports the same premise: fusion reduces mean summed Brier loss by 33.7\% and 35.1\% relative to the classical and quantum views alone. Retrospective validation-based selection distinguishes these regimes by choosing fusion for the mixed targets and single views for the pure targets, while replacement of the target-generating circuit with an independent circuit removes the large constructed-task gain. Together, these controls link the observed benefit to the specific representation and target relationship. The block-target tests further show why the fusion rule must match additive or interaction relationships.

We presented a classical--quantum multi-view framework that retains the full classical input and combines it with circuit-derived features through concatenation or attention. Benchmark validation shows improved mean macro-$F_1$ on Wine, Fashion-MNIST and Steel Plates Faults and the highest WDBC mean within the matched calibrated comparison. Classical baselines lead on Digits and Covertype, matched classical controls remain competitive, and native-view outcomes vary with the available inputs. Together, these findings establish useful gains under the evaluated conditions and support matching the representation and fusion strategy to the target relationships. Treating classical and quantum representations as complementary views, with controls that assess their contributions, offers a systematic route to identifying tasks where quantum representations add predictive value.

\declarationheading{Data and code availability}
Public dataset identifiers, evaluated subsets and source limitations are documented in Supplementary Section~S1. The controlled data-generating process, sample sizes and seeds are specified in Supplementary Section~S4. The supplement reports the training settings and retained experimental summaries, including the five-benchmark results in Section~S7. Section~S8 specifies the angular--polynomial diagnostic, for which the accompanying reproducibility files provide the generator, fitting and verification code, per-repetition outputs and complete summaries. 
The code is available via \url{https://github.com/azalya-green/Quantum_Classical_MultiView}

\declarationheading{CRediT authorship contribution statement}
\textbf{Azadeh Alavi, Fatemeh Kouchmeshki and Hossein Akhoundi:} Conceptualization, Methodology, Software, Writing -- original draft, Writing -- review \& editing. \textbf{Yongli Ren, Abdolrahman Alavi and Muhammad Usman:} Writing -- review \& editing (manuscript comments and editing).

\declarationheading{Funding}
This research did not receive any specific grant from funding agencies in the public, commercial, or not-for-profit sectors.

\declarationheading{Declaration of competing interests}
This research was undertaken as a project of Pattern Recognition Pty Ltd, which has a commercial interest in the research software. Azadeh Alavi, Fatemeh Kouchmeshki and Hossein Akhoundi are named applicants on Australian provisional patent application 2026906063, filed on 6 July 2026. The application concerns geometry-aware quantum workload selection, circuit replacement and circuit synthesis.

\declarationheading{Generative AI}
During the preparation of this work, the authors used ChatGPT and Claude for editing. The authors reviewed and edited the output as needed and take full responsibility for the content of the published article.

\par\doublespacing\normalsize
\setlength{\bibsep}{0pt}
\bibliographystyle{elsarticle-num}
\bibliography{references}
\end{document}